\documentclass[10pt, paper=a4, UKenglish]{article}
\usepackage{graphicx}
\def\Title#1{\begin{center} {\Large #1 } \end{center}}
\def\Author#1{\begin{center}{ \sc #1} \end{center}}
\def\Address#1{\begin{center}{ \it #1} \end{center}}

\newcommand\pubblock{\rightline{\begin{tabular}{l} Proceedings of the CTD/WIT 2019\\ \pubnumber\\
         \pubdate  \end{tabular}}}

\newenvironment{Abstract}{\begin{quotation} \begin{center} 
             \large ABSTRACT \end{center}\bigskip 
      \begin{center}\begin{large}}{\end{large}\end{center} \end{quotation}}

\newenvironment{Presented}{\begin{quotation} \begin{center} 
             PRESENTED AT\end{center}\bigskip 
      \begin{center}\begin{large}}{\end{large}\end{center} \end{quotation}}

\def\beq{\begin{equation}}
\def\eeq#1{\label{#1}\end{equation}}
\def\eeqn{\end{equation}}

\def\beqa{\begin{eqnarray}}
\def\eeqa#1{\label{#1}\end{eqnarray}}
\def\eeqan{\end{eqnarray}}

\let\bar=\overbar

\def\Dslash{\not{\hbox{\kern-4pt $D$}}}
\def\dslash{\not{\hbox{\kern-2pt $\del$}}}

\def\msb{{\bar{\ssstyle M \kern -1pt S}}}

\usepackage{color}
\usepackage{lineno}
\usepackage{subfig}
\usepackage{hyperref}

\newcommand\pubnumber{PROC-CTD19-134}

\newcommand\pubdate{\today}

\def\affiliation{
Instituto de F\'isica Corpuscular (CSIC-UV), Spain}

\newcommand{\conference}{Connecting the Dots and Workshop on Intelligent Trackers (CTD/WIT 2019)\\
Instituto de F\'isica Corpuscular (IFIC), Valencia, Spain\\ 
April 2-5, 2019}

\usepackage{fancyhdr}
\definecolor{mygrey}{RGB}{105,105,105}
\begin{document}


\large
\begin{titlepage}
\pubblock

\vfill
\Title{Overview of CMOS sensors for future tracking detectors}
\vfill

\Author{Ricardo Marco Hern\'{a}ndez}
\Address{\affiliation}
\vfill

\begin{Abstract}
The depleted CMOS sensors are emerging as one of the main candidate technologies for future tracking detectors in high luminosity colliders. Its capability of integrating the sensing diode into the CMOS wafer hosting the front-end electronics allows for reduced noise and higher signal sensitivity. They are suitable for high radiation environments due to the possibility of applying high depletion voltage and the availability of relatively high resistivity substrates. The use of a CMOS commercial fabrication process leads to their cost reduction and allows faster construction of large area detectors. A general perspective of the state of the art of these devices will be given in this contribution as well as a summary of the main developments carried out with regard to these devices in the framework of the CERN RD50 collaboration.
\end{Abstract}

\vfill

\begin{Presented}
\conference
\end{Presented}
\vfill
\end{titlepage}
\def\thefootnote{\fnsymbol{footnote}}
\setcounter{footnote}{0}
%

\normalsize 


\section{Introduction}
\label{intro}

\begin{figure}[!htb]
	\centering
	\includegraphics[width=0.25\linewidth]{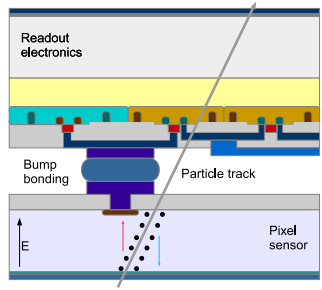}
	\caption{Diagram of a hybrid pixel detector.}
	\label{fig:hybrid_pixel}
\end{figure}

Current large pixel detectors in High Energy Physics, such as the ones which will be part of the ATLAS Tracker Detector \cite{atlas} or the CMS Tracker Detector \cite{cms} upgrades for the High Luminosity Large Hadron Collider (HL-LHC), mostly follow a hybrid aproach. In a hybrid pixel detector the sensor and the readout electronics are independent devices connected by means of bump-bonding, as it can be seen in Figure~\ref{fig:hybrid_pixel}. This fact allows for independent development of the sensor and readout electronics technologies to cope with high radiation environments and particle rates. However, bump-bonding is a complex and expensive assembly process with a limited output rate. Moreover, the overall material budget of several layers restricts the accuracy of the particle trajectory measurement in hybrid pixel detectors.

\begin{figure}[!htb]
	\centering
	\includegraphics[width=0.5\linewidth]{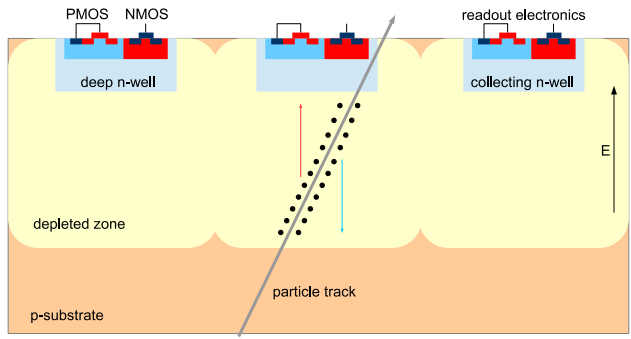}
	\caption{Diagram of a depleted CMOS MAPS sensor.}
	\label{fig:cmos_det}
\end{figure}

A promising alternative to the current hybrid approach is the so-called depleted monolithic active pixel sensors (MAPS). These kind of sensors integrate the sensing diode and the readout electronics in the same CMOS wafer, as it is shown in Figure~\ref{fig:cmos_det}. The charge is mainly collected in the depleted area created by applying reverse bias. The pixel electronics is placed inside an isolated n-well. This sensor technology offers the possibility of noise reduction and higher signal sensitivity. An important adavntage is that its commercial CMOS fabrication process and its integration leads to an easier production, a large cost reduction and a faster fabrication. Furhtermore, this technology can save material budget due to its reduced thickness.

Unfortunately, although the depleted CMOS sensor technology offers several clear advantages, there are several aspects of this technology which still have to be improved, as the radiation tolerance, the timing resolution and a fast readout capbility to cope with high particle rates.

\section{Large versus small fill-factor structures}
\label{section}

\begin{figure}[!htb]
	\centering
	\subfloat[]{\includegraphics[width=0.4\linewidth]{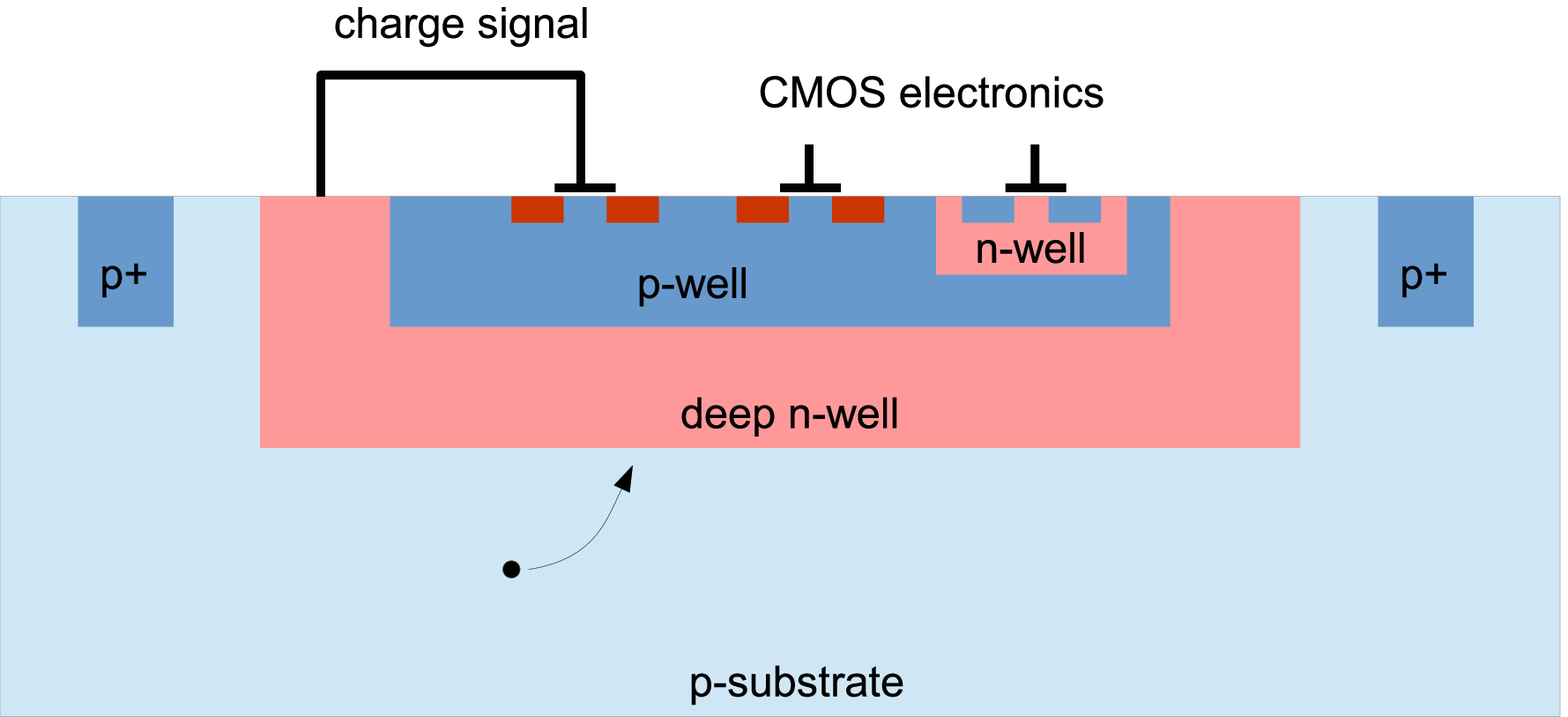}}
	\qquad
	\subfloat[]{\includegraphics[width=0.4\linewidth]{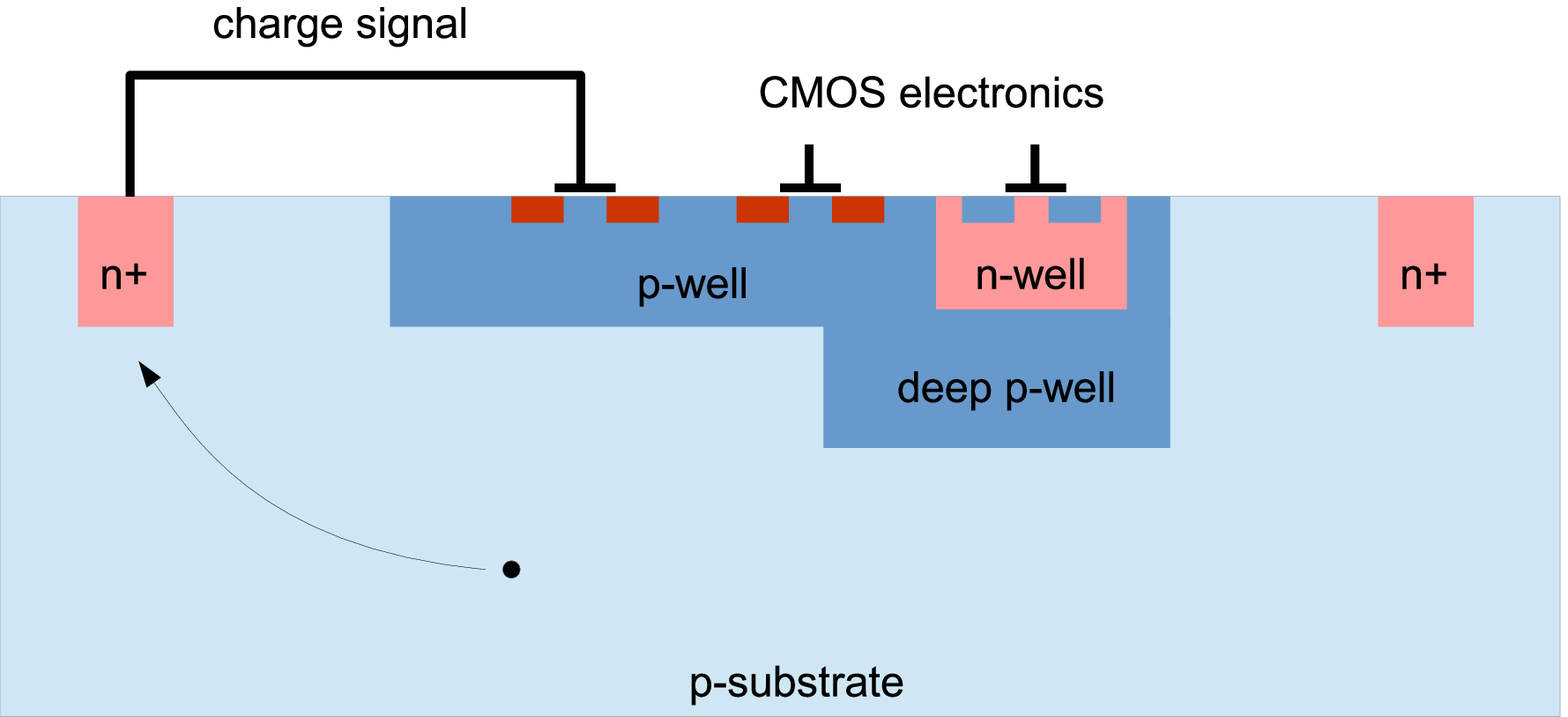}}
	\caption{Large fill-factor structure (a) and small fill-factor structure (b).}
	\label{fig:cmos_struct}
\end{figure}

The two main n-on-p CMOS sensor design concepts according to their collection electrode size, known as the fill factor, are depicted in Figures~\ref{fig:cmos_struct}(a--b). In the large-fill factor structure (Figure~\ref{fig:cmos_struct}a), the sensing diode is made up of the p-substrate and the deep n-well while the electronics are place,d inside the charge collection well. High resisitvity substrates, up to 3 k$\Omega$$\cdot$cm \cite{wermes}, are currently available for this structure. Since a high bias voltage can be applied to the device, there will be on average shorter drift distances and, as a consequence, a higher radiation tolerance capability. However, the sensor capacitance is larger in this structure, up to hundreds of fF \cite{hemperek1} depending on the pixel size, so the noise will be also larger. The electronics will also have lower speed and will need more power to counterbalance this fact. Furthermore, this structure is more prone to cross talk from digital electronics into sensor.

On the other hand, the small fill-factor structure (Figure~\ref{fig:cmos_struct}b) is characterized by the fact that the sensor diode and the readout electronics are separated by the p-substrate. Higher resistivity substrates, up to 8 k$\Omega$$\cdot$cm \cite{wermes}, can be used with this structure but only low bias voltage can be applied to the substrate. Therefore, longer drift distances are required on average and consequently, this structure tends to a lower radiation tolerance. Nevertheless, the sensor capacitance is very small, a few fF \cite{hemperek1},  having a lower noise compared to the large fill-factor structure. Thus, the electronics implemented in the small fill-factor structure can be faster and less power-consuming.

\section{Challenges of the depleted CMOS sensors}
\label{section}

As it has been aforementioned, there are three important challenges that the depleted CMOS sensor technology has to face: the radiation hardness, the timing resolution and a fast data readout.

\subsection{Radiation tolerance}
\label{subsection}

The first challenge is to increase the radiation hardness of this sensor technology, currently about 10$^{15}$ 1 MeV n$_{eq}$/cm$^{2}$ \cite{vilella1}, to reach the equivalent fluences required, for instance, for the Future Circular Collider, which will be larger than 7$\cdot$10$^{17}$ 1 MeV n$_{eq}$/cm$^{2}$ \cite{besana}. In order to meet this requirement, the CMOS sensors will need to increase the allowed substrate biasing as well as to access to high resistivity substrates, since the depleted depth is proportional to the product of these two parameters. In that sense, the effective doping concentration of the p-substrate varies with irradiation in a different way for high and low resistivities \cite{mandic1}, there is a increase of the depletion depth after irradiation for lower resistivities (up to equivalent fluences of about 2$\cdot$10$^{15}$ 1 MeV n$_{eq}$/cm$^{2}$) whereas the depletion depth decreases after irradiation for higher resistivities. This fact has to be considered in the sensor design process. The backside processing of CMOS sensors is also an improtant factor to increase their radiation tolerance since having back bias contact or thinned devices will improve their charge collection efficiency. Another relevant aspect for a better radiation hardness would be the possibility of implementing multiple nested wells in order to increase the isolation between the CMOS electronics and the substrate so that higher bias voltages could be applied to the devices. Finally, the use of CMOS technologies with smaller features sizes would also increase the radiation tolerance of the depleted CMOS sensors.

\subsection{Timing resolution}
\label{subsection}

The second challenge is to improve the timing resolution of this kind of devices from the current one, lower than 10 ns \cite{augustin}\cite{benoit}, to a even better resolution, lower than 5 ns. There are different sources of time uncertainty such as the charge collection time, the delay in the readout electronics and the time-walk in the comparator. The reduction of the charge collection time is constrained by the sensor geometry and bias whereas the electronics delay improvement is limited by the readout electronics power consumption. Therefore, there is more room for improvement in the reduction of the time-walk by applying new design methods for the discrimination of the analogue signals like using a time-walk compensated comparator \cite{vilella1}, the two-threshold method or the ramp method \cite{augustin}.

\subsection{Fast readout}
\label{subsection}

The third challenge for this technology is to be able to get a fast readout of the data to deal with high particle rates in future high luminosity colliders. For instance, a particle rate higher than 1000 MHz/cm$^{2}$ is foreseen in the inner pixels layers of the ATLAS tracker detector in the HL-LHC \cite{atlas}. One of the advantages of the depleted CMOS sensor technology is the implementation of the full readout architecture in the same device to cope with detector demands, not only in terms of particle rates but also for triggering or time stamping. There are different readout architectures such as the column drain architecture implemented in the LF-Monopix device \cite{wang} or the parallel pixel to buffer architecture implemented in the ATLASPix M2 device \cite{kiehn}. In the former architecture, the address and time stamp is read out in each pixel and then the data are moved selectively to the chip periphery whereas in the latter architecture the binary data of each pixel are moved inmediately to the device periphery and proccessed there upon a trigger arrival. Another type of asynchronous readout architecture has been also implemented in the TJ-Malta device \cite{argemi}. These readout architectures are being tested to optimize the integration, cross-talk and speed of the device.

\section{Commercial CMOS foundries}
\label{section}

There is a number of commercial foundries available for the fabrication of depleted CMOS sensors. The CMOS sensors community has already a valuable experience with some of these vendors from several developments carried out. Large fill-factor structures have been produced in LFoundry Srl, ams AG and TSI Semiconductors Corp, in the framework of the ATLAS upgrade and the Mu3e experiments, like the H35Demo device \cite{vilella2}, implemented in the ams 350 nm process , the LF-Monopix device \cite{wang}, implemented in the LFoundry 150 nm process, or the MuPix7 and MuPix8 devices \cite{augustin}, implemented in the ams and TSI 180 nm processes, to mention a few ones. Regarding the small fill factor structures, some devices of this type have been also produced in TowerJazz Ltd. following the TJ 180 nm process as the TJ-Monopix device \cite{wang} or the TJ-Malta device \cite{argemi}. A special type of CMOS technology, the High Voltage Silicon On Insulator, is also available at X-FAB Semiconductor Foundries AG, several devices have been implemented in the XFAB 180 nm process, like the XTB01 device \cite{hemperek2}. Finally, it must be emphasized that the current technologies available for CMOS sensors offer feature sizes larger than 130 nm, therefore the radiation tolerance and logic dencity of these devices can be further improved with smaller sizes in the future.

\section{Summary of CMOS activities in the framework of the CERN RD50 collaboration}
\label{section}

The CERN RD50 is an international collaboration with more than 300 members aimed at developing and characterizing radiation-hard semiconductor devices for high luminosity colliders. As it has been mentioned, semiconductor sensors will be exposed in the HL-LHC or FCC to hadron fluences not withstood by current LHC sensors. Among other research interests, the collaboration has a research line in new detector structures, such as n-on-p sensors, 3D sensors, low gain avalanche photodiodes and depleted CMOS sensors. The latter are a priority for RD50. In fact, several depletion depth and charge collection measurements have been already carried out with different CMOS devices \cite{mandic2}\cite{caravallo}. Moreover, there is a new effort within the RD50 collaboration to develop different matrices of pixels and test structures in depleted CMOS processes.

\begin{figure}[!htb]
	\centering
	\includegraphics[width=0.5\linewidth]{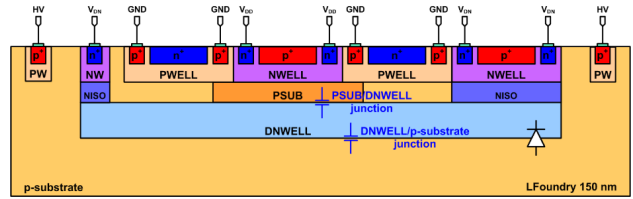}
	\caption{Diagram of the RD50-MPW1 depleted CMOS sensor cross-section.}
	\label{fig:rd50_mpw1_cs}
\end{figure}

A small depleted CMOS sensor prototype, the RD50-MPW1 \cite{vilella3} device of 5 mm by 5 mm, has been already developed using the 150 nm LFoundry process in two different resistivities, 500 $\Omega$$\cdot$cm and 1.9 k$\Omega$$\cdot$cm. The device has a total thickness of 280 µm. The main goals of this design are to test the technology used and verify the validity of novel designs. Figure~\ref{fig:rd50_mpw1_cs} shows a diagram of the RD50-MPW1 device cross-section, where it can be seen that the device has a large fill factor structure. The RD50-MPW1 device has test structures for Edge-Transient Current Technique (E-TCT) measurements and two independent CMOS pixels matrices. One is a photon counting matrix with 28 by 52 pixels with embedded readout electronics, charge amplifier and discriminator, and a 16-bit counter. The other matrix has 40 by 78 pixels, with a size of 50 $\mu$m by 50 $\mu$m, with embedded readout electronics following a column drain readout architecture similar to the FEI3 readout chip \cite{peric}.  The DAQ development and the device characterization activities related to the RD50-MPW1 device are still ongoing. The sensors are fully functional but the measured leakage current is higher than expected.

\begin{figure}[!htb]
	\centering
	\includegraphics[width=0.5\linewidth]{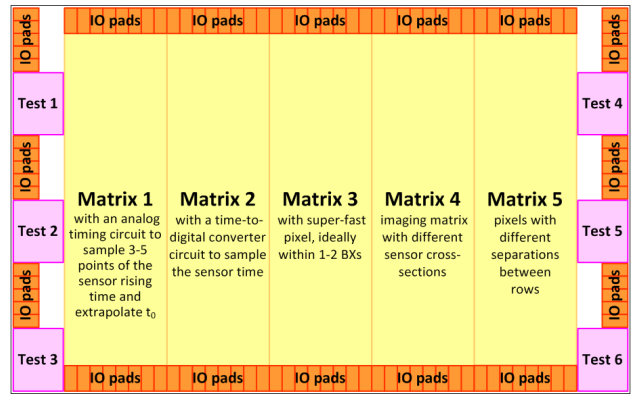}
	\caption{Diagram of the RD50-ENGRUN1 device main blocks.}
	\label{fig:rd50_engrun1}
\end{figure}

Finally, the so-called RD50-ENGRUN1 device, a large area demonstrator, is also being designed within the CERN RD50 collaboration using the 150 nm process of Lfoundry. A diagram of the main blocks of the RD50-ENGRUN1 device can be seen in Figure~\ref{fig:rd50_engrun1}. This device has several independent depleted CMOS pixel matrices and test structures. The pixel matrices have different purposes and the main goals pursued in this design are the improvement of the current time resolution with dedicated readout circuits, the implementation of new sensor cross-sections, the assesment of pre-stitching options to incrase the device size beyond the reticle size limitation and the increase of radiation tolerance by sensor design and backside processing. The design effort is being carried out by several of the institutions already involved in the RD50 CMOS project. The DAQ development and TCAD simulations are also running in parallel with the device design.

\section{Conclusions}

It has been pointed out that the depleted CMOS sensor technology is very promising for future silicon tracking detectors due to its lower cost, easier and faster detector assembly, faster fabrication turn-around and larger material budget saving. However, some challenges must still be faced to cope with future tracking detectors requirements in terms of radiation hardness, timing resolution and fast data readout. This sensor technology is a priority for the CERN RD50 collaboration, which has been involved in the radiation tolerance study of depleted CMOS sensors. Within the CERN RD50 collaboration, a project to develop depleted CMOS sensors has been started. The RD50-MPW1 chip has been designed and fabricated ans is currently under test. The RD50-MPW2 has been designed and already submitted aimed at reducing the leakage current and improving the readout electronics speed. A larger RRD50-ENGRUN1 device is being designed with several matrices of pixels with the main goal of improving the timing resolution.






\begin{thebibliography}{99}


\bibitem{atlas} 
ATLAS Collaboration, "Technical Design Report for the ATLAS Inner Tracker Pixel Detector", CERN-LHCC-2017-021, September 2017.

\bibitem{cms} 
CMS Collaboration, "The Phase-2 Upgrade of the CMS Tracker (Technical Design Report)", CERN-LHCC-2017-009, June 2017.

\bibitem{wermes} 
Wermes, N., "CMOS pixel sensors/detectors overwiew", 32nd RD50 Workshop, Hamburg (Germany), 4-6 June 2018.

\bibitem{hemperek1} 
Hemperek, T., "Overview and perspectives of depleted CMOS sensors for high radiation environments", Proceedings of Science (Vertex 2017) {\bf 034}.

\bibitem{vilella1} 
Vilella, E. et al., "Radiation tolerance and time resolution of depleted CMOS sensors", Proceedings of Science (Vertex 2018) {\bf In Press}.

\bibitem{besana} 
Besana, M. I. et al., "Evaluation of the radiation field in the future circular collider detector", Phys. Rev. Accel. Beams {\bf 19} (2016) 111004. DOI: 10.1103/PhysRevAccelBeams.19.111004.

\bibitem{mandic1} 
Mandic, I. et al., "Neutron irradiation test of depleted CMOS pixel detector prototypes", JINST {\bf 12} (2017) P02021. DOI:10.1088/1748-0221/12/02/P02021.

\bibitem{augustin} 
Augustin, H. et al., "Performance of MuPix8 a large scale HV-CMOS pixel sensor", International Workshop on Semiconductor Pixel Detectors for Particles and Imaging (PIXEL 2018), Taipei (Taiwan), 10-14 December 2018.

\bibitem{benoit} 
Benoit, M., "Characterization of the 180nm HV-CMOS ATLASPix large-fill factor Monolithic prototype", 14th Trento Workshop on Advanced Silicon Radiation Detectors, Trento (Italy), 25-27 February 2019.

\bibitem{wang} 
Wang, T. et al., "Depleted fully monolithic CMOS pixel detectors using a column based readout architecture for the ATLAS Inner Tracker upgrade", JINST {\bf 13} (2018) C03039. DOI:10.1088/1748-0221/13/03/C03039.

\bibitem{kiehn} 
Kiehn, M. et al., "Performance of CMOS pixel sensor prototypes in ams H35 and aH18 technology for the ATLAS ITk upgrade", Nuclear Inst. and Methods in Physics Research, A {\bf 924} 104–107 (2019). DOI:10.1016/j.nima.2018.07.061.

\bibitem{argemi} 
Argemi, L. et al., "Development of the monolithic "MALTA" CMOS sensor for the ATLAS ITk outer pixel layer", Proceedings of Science (TWEPP 2018) {\bf In Press}.

\bibitem{vilella2} 
Vilella, E. et al., "Prototyping of an HV-CMOS demonstrator for the High Luminosity-LHC upgrade", JINST {\bf 11} (2016) C01012. DOI:10.1088/1748-0221/11/01/C01012.

\bibitem{hemperek2} 
Hemperek, T. et al., "A Monolithic Active Pixel Sensor for ionizing radiation using 180 nm HV-SOI process", Nuclear Inst. and Methods in Physics Research, A {\bf 796} 8–12 (2015). DOI: 10.1016/j.nima.2015.02.052.

\bibitem{mandic2} 
Mandic, I. et al., "Charge-collection properties of irradiated depleted CMOS pixel test structures", Nuclear Inst. and Methods in Physics Research, A {\bf 903} 126–133 (2018). DOI:10.1016/j.nima.2018.06.062.

\bibitem{caravallo} 
Caravallo, E. et al., "Studies of irradiated AMS H35 CMOS detectorsfor the ATLAS tracker upgrade", JINST {\bf 12} (2017) C01074. DOI:10.1088/1748-0221/12/01/C01074.

\bibitem{vilella3} 
Vilella, E. et al., "Overview of design and evaluation of depleted CMOS sensors within RD50", 3rd RD50 Workshop, CERN (Switzerland), 26-28 November 2018.

\bibitem{peric} 
Peric, I. et al., "The FEI3 readout chip for the ATLAS pixel detector", Nuclear Inst. and Methods in Physics Research, A {\bf 565} 178-187 (2006). DOI: 10.1016/j.nima.2006.05.032.

\end{thebibliography}
\end{document}